\documentclass[letterpaper,twocolumn,english,superscriptaddress,prl,aps,showpacs]{revtex4}
\usepackage[T1]{fontenc}
\usepackage[latin1]{inputenc}
\usepackage{babel}
\usepackage{amsmath}
\usepackage{amssymb}
\usepackage{graphicx}

\makeatletter

\pdfpageheight\paperheight
\pdfpagewidth\paperwidth

\@ifundefined{textcolor}{}
{%
 \definecolor{BLACK}{gray}{0}
 \definecolor{WHITE}{gray}{1}
 \definecolor{RED}{rgb}{1,0,0}
 \definecolor{GREEN}{rgb}{0,1,0}
 \definecolor{BLUE}{rgb}{0,0,1}
 \definecolor{CYAN}{cmyk}{1,0,0,0}
 \definecolor{MAGENTA}{cmyk}{0,1,0,0}
 \definecolor{YELLOW}{cmyk}{0,0,1,0}
}

\@ifundefined{definecolor}{\@ifundefined{definecolor}
 {\@ifundefined{definecolor}
 {\@ifundefined{definecolor}
 {\usepackage{color}}{}
}{}
}{}
}{

}\usepackage{babel}
\DeclareMathAlphabet{\mathpzc}{OT1}{pzc}{m}{it}

\makeatother

\begin{document}

\author{Alexey A. Kovalev}

\affiliation{Department of Physics \& Astronomy, University of California, Riverside,
California 92521, USA}

\affiliation{Department of Physics \& Astronomy, University of Nebraska-Lincoln,
Lincoln, NE 68588, USA}

\author{Amrit De}

\affiliation{Department of Physics \& Astronomy, University of California, Riverside,
California 92521, USA}

\author{Kirill Shtengel}

\affiliation{Department of Physics \& Astronomy, University of California, Riverside,
California 92521, USA}

\title{Spin Transfer of Quantum Information between Majorana Modes and a Resonator}

\date{\today}
\begin{abstract}
We show that resonant coupling and entanglement between a mechanical
resonator and majorana bound states can be achieved via spin currents
in a 1D quantum wire with strong spin-orbit interactions. The bound
states induced by vibrating and stationary magnets can hybridize thus
resulting in spin-current induced $4\pi$-periodic torque, as a function
of the relative field angle, acting on the resonator. We study the
feasibility of detecting and manipulating majorana bound states with
the use of magnetic resonance force microscopy techniques.
\end{abstract}
\maketitle
\textit{Introduction.} --- Majorana zero states bound to domain walls
in 1D and quasi-1D systems such as $p$-wave superconducting wires
~\cite{Kitaev:2001}, edges of 2D topological insulators~\cite{Fu:2008,Nilsson:Sep2008}
and semiconducting quantum wires with strong spin-orbit interactions~\cite{Lutchyn:2010,Oreg:2010}
can be potentially utilized to form non-local qubits thus providing
a platform for topological quantum computing~\cite{Nayak:Sep2008,Alicea:2012,Beenakker:2013}.
Of these systems, spin-orbit-coupled semiconductor wires with proximity-induced
superconductivity are of particular practical interest, with a number
of recent experiments aiming at establishing the existence of Majorana
bound states (MBS) there~\cite{Mourik:Science2012,Deng:NanoLett2012,Das:2012,Rokhinson:nov2012}.
While further studies are needed to unambiguously confirm their existence~\cite{Bagrets:Nov2012,Liu:Dec2012,Pikulin:2012,Lee:Oct2012,DasSarma:2012,Finck:Mar2013},
one can also look ahead and try developing efficient techniques for
manipulating MBS~\cite{Alicea:Nat2011,Teo:Jan2010,Sau:Nov2010,Sau:Sep2011,Flensberg:Mar2011,Halperin:Apr2012,Romito:PRB2012}.
Several recent proposals addressed the possibility of control of topological
qubits by coupling them to more conventional ones, such as flux qubits
via the Aharonov--Casher effect~\cite{Hassler:2010,Jiang:prl2011,Bonderson:prl2011,Hassler:2011,Pekker:unpub}.

Proposals related to observation of MBS quite often rely on tunneling and
transport effects that are indicative of the zero energy nature of these
modes
\cite{Bagrets:Nov2012,Liu:Dec2012,Pikulin:2012,Lee:Oct2012,DasSarma:2012,Finck:Mar2013}.
Some recent proposals are also related to unconventional Josephson effect in
Majorana quantum wires and TI edges where the periodicity is equal to $4\pi$
\cite{Kitaev:2001,Fu:Apr2009,Lutchyn:2010,Jiang:Nov2011}. A dual effect
whereby a torque between magnets exhibits $4\pi$ periodicity in the field
orientations has also been
suggested~\cite{Meng:PRB2012,Jiang:prb2013,Kotetes:julunpublished}. It is
this effect that can lead to mechanical torques and quantum information
transfer between MBS and a mechanical resonator.
The idea of coupling a two-level system to vibrational modes to form a hybrid quantum system has been successfully used in quantum optics \cite{Groblacher:Nat2009} and, more recently, in the field of nanomechanical resonators where a single phonon control has been demonstrated \cite{O'Connell:apr2010}. We propose using a similar technique in the context of topological qubits.

\begin{figure}
\includegraphics[clip,width=0.4\textwidth]{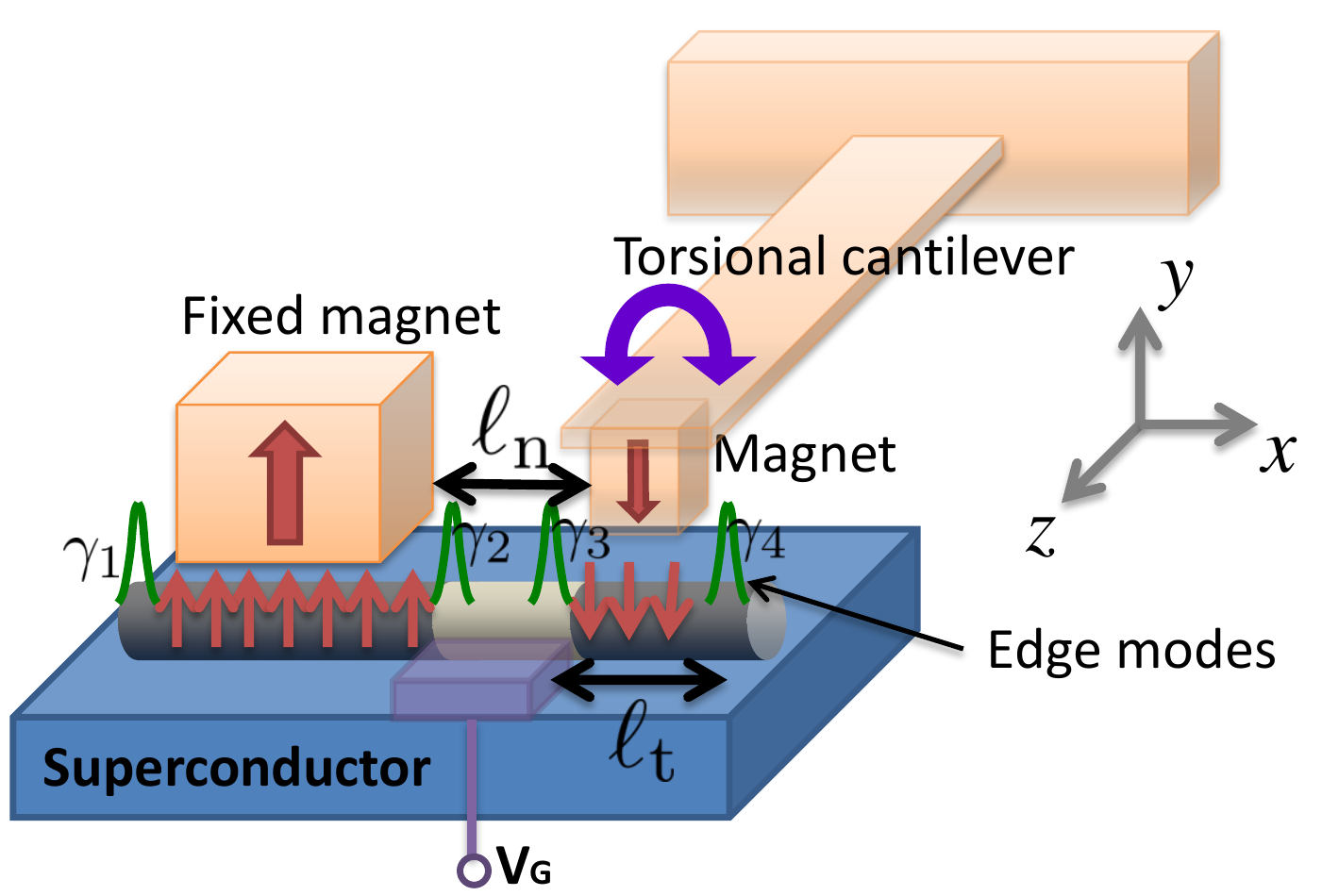} \caption{(Color online) A 1D semiconductor wire with strong spin orbit interaction
is placed on top of s-wave superconductor. Majorana bound states are
defined by magnetic fields of two magnets, one of which is free to
vibrate. The gate $V_{G}$ can be used in order to control the hybridization. }

\label{fig1}
\end{figure}

It has been predicted that
conservation of angular momentum in macrospin molecules can result in quantum
entanglement of a tunneling spin with mechanical modes
\cite{Kovalev:PRL2011,Garanin:PRX2011}. A flow of spin current between two
magnets has been demonstrated to induce spin-transfer torque effect
\cite{Berger1996,Slonczewski1996} and mechanical torques
\cite{Kovalev:Jan2007,Zolfagharkhani:dec2008}, also by conservation of
angular momentum.
In this Letter, we study resonant coupling between a Majorana qubit and a
mechanical resonator induced by spin currents flowing over portions (region
of length $\ell_{\text{n}}$ in Fig.~\ref{fig1}) of 1D semiconductor quantum
wire. This resonant coupling is controlled by non-dissipative spin currents
in a spin-transistor type architecture \cite{Datta:1990} -- which effectively
allows or disallows the hybridization of two MBS. A nano-magnet attached to
the resonator then feels the hybridization as a mechanical torque which can
result in the state (quantum information) transfer between the Majorana qubit
and the mechanical resonator.

A Majorana qubit is formed by four MBS where three of these MBS are
hybridized (Fig.~\ref{fig1}). The non-topological region could be formed by
magnets with sharp field profiles or by hetero-junction nanowires with
contrasting $g-$factors. The effective low energy Hamiltonian then becomes:
\begin{equation}
\mathcal{H}=\hbar\omega_{\text{r}}a^{\dagger}a+iE^{\text{n}}(\theta)\gamma_{2}\gamma_{3}+iE^{\text{t}}\gamma_{3}\gamma_{4}\,,\label{eq:Ham1}
\end{equation}
where $a$ is the annihilation operator of the resonant torsional mode of the
cantilever so that $\theta=\theta_{0}+\theta_{\text{zpf}}(a^{\dagger}+a)$
with $\theta_{\text{zpf}}=(\hbar^{2}/KI)^{1/4}$ being the angle of zero point
fluctuations of the cantilever, $K$ is the spring constant, $I$ is the moment
of inertia, $E^{\text{t(n)}}$ describes the hybridization energy,
$\gamma_{i}$ describes MBS. It is the $\theta$ dependence of the
hybridization energy that leads to three interrelated effects: (i) coupling
of the rotation of the magnet to the internal state of Majorana qubit, (ii)
mechanical torque acting on the magnets and (iii) spin current
$j_{s}^{z}(x)=\mathfrak{Re}\left[\Psi^{\dagger}(x)\hat{\sigma}^{z}\hat{\upsilon}\Psi(x)\right]$
defined in the non-topological middle section in which the magnetic field is
absent, here the velocity operator is
$\hat{\upsilon}=\partial\hat{H}/\partial p$. We obtain that the
torque on the magnets
\cite{Meng:PRB2012,Jiang:prb2013,Kotetes:julunpublished} is generated
solely by the spin current passing through the middle non-topological region when there is no hybridization over the topological regions in Fig
\ref{fig1}.

According to our estimates, strong coupling between the Majorana qubit and
the mechanical resonator can lead to a shift in the mechanical resonant
frequency, Rabi oscillations, coherent state transfer and entanglement. All
these effects could signify a presence of a Majorana qubit. This mechanism
can also be utilized to couple several Majorana qubits or to couple a
Majorana qubit with a non-topological qubit such as an NV center
\cite{Kolkowitz:Science2012}.

\textit{Spin currents and edge hybridization.} --- We consider a
semiconductor wire with strong spin-orbit interaction in the presence of a
Zeeman field (note that a TI edge gives qualitatively similar results). The
wire is proximity-coupled to an $s$-wave superconductor which induces the
pairing strength $\Delta$ in the wire. A topological region is induced by
external magnets (Fig.~\ref{fig1}) where one of the magnets is attached to a
mechanical resonator and can mechanically vibrate at frequency
$\omega_{\text{r}}\ll\Delta$.

The 1D wire is described by a BdG Hamiltonian:
\begin{equation}
\begin{aligned}\hat{H}= & p^{2}\hat{\tau}^{z}/2m^{*}+\alpha_{\text{so}}k\hat{\tau}^{z}\hat{\sigma}^{z}-\mu\hat{\tau}^{z}+\Delta(\cos\phi\hat{\tau}^{x}-\sin\phi\hat{\tau}^{y})\\
 & -b\hat{\sigma}^{z}+B(\cos\theta\hat{\sigma}^{x}-\sin\theta\hat{\sigma}^{y})\,,
\end{aligned}
\label{eq:BdG}
\end{equation}
where $m^{*}$ is the effective mass, $\alpha_{\text{so}}$ is the strength of
spin-orbit interaction, $\mu$ is the chemical potential, $\Delta e^{i\phi}$
is the superconducting pairing, $b$ is the magnetic field along the
$z$-direction and $B$ is the magnetic field in the $xy$-plane. Here we use
the Nambu spinor basis
$\Psi^{T}=(\psi_{\uparrow},\psi_{\downarrow},\psi_{\downarrow}^{\dagger},-\psi_{\uparrow}^{\dagger})$
and the Pauli matrices $\hat{\sigma}^{i}$ and $\hat{\tau}^{i}$ describe the
spin and particle-hole sectors, respectively.

The Hamiltonian~(\ref{eq:BdG}) supports both gapped and gapless phases, its
phase diagram is more complicated compared to the TI edge
system~\cite{Meng:PRB2012,Jiang:prb2013} whose Hamiltonian does not contain
the $p^{2}\tau^{z}$ term. Here we restrict ourselves to the case
$\Delta^{2}>b^{2}$ so that the Hamiltonian~(\ref{eq:BdG}) describes two
gapped phases: topological (T) if $\Delta^{2}-b^{2}<B^{2}-\mu^{2}$ and
non-topological (N) if $\Delta^{2}-b^{2}>B^{2}-\mu^{2}$, separated by a
quantum phase transition at $\Delta^{2}-b^{2}=B^{2}-\mu^{2}$.
\begin{figure}
\includegraphics[clip,width=0.48\textwidth]{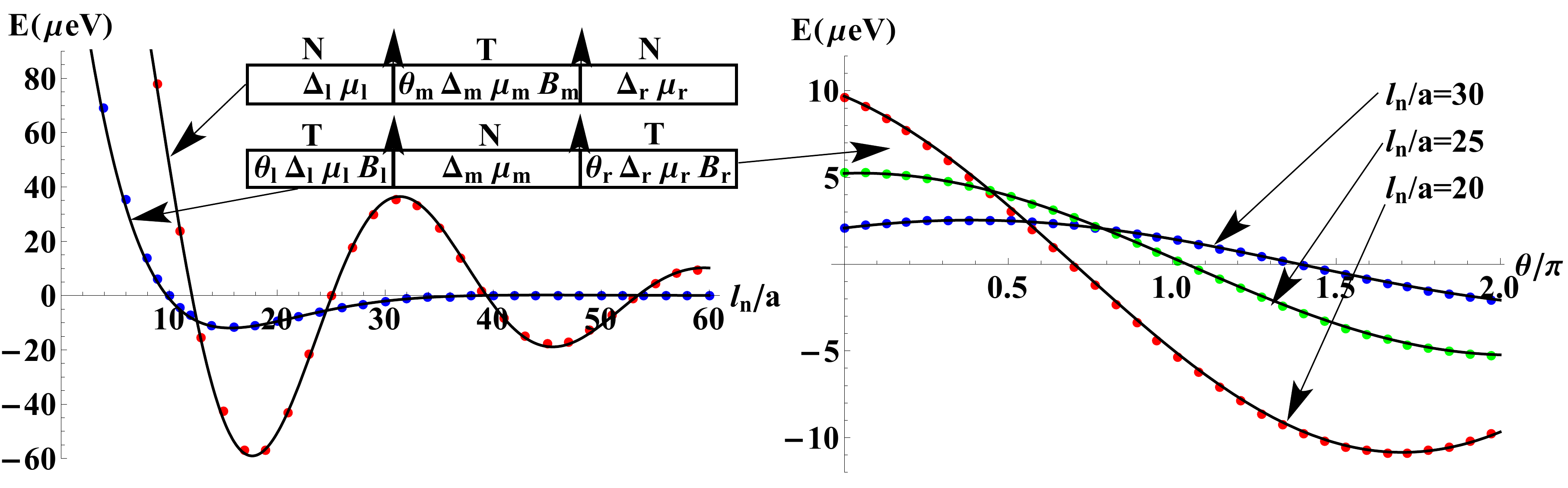} \caption{(Color online) Hybridization energies of two Majorana bound states over
topological and non-topological regions in a semiconductor wire as a function of the hybridization region length (left) and
relative angle of magnetic fields (right), only for T--N--T structures. The circles represent the corresponding numerical results.}

\label{fig2}
\end{figure}

We analyze analytically hybridization of the edge modes which results in spin
currents and torques in N--T--N and T--N--T setups shown in Fig.~\ref{fig2} where
we have an infinite semiconductor wire with a finite topological (T) or
non-topological (N) region. We assume that the phase of the superconducting
pairing is constant throughout the wire, the magnetic field is always zero
for N-regions and $b=0$ in all regions. Then gapped regions are described by
parameters $\{\Delta,B,\mu,\theta\}$ for the T-region and by $\{\Delta,\mu\}$
for the N-region (see Figs.~\ref{fig2}). We first determine the bound state
of a single T--N boundary by finding $4-$component zero energy solution to the
Hamiltonian~(\ref{eq:BdG}) in the form $\Psi(x)=e^{\kappa x}\Psi(\kappa)$. In
general, we arrive at four solutions that decay into the topological region,
i.e. with $\mathfrak{Re}(\kappa)>0$, and four solutions that decay into the
non-topological region, i.e. with $\mathfrak{Re}(\kappa)<0$. A linear
combination of these solutions on each side has to be continuous and have a
continuous derivative at the boundary between T and N-regions leading to a
unique solution for MBS. We denote such normalized solutions as
$\left|\psi_{\text{L}}\right\rangle $ for the left Majorana and as
$\left|\psi_{\text{R}}\right\rangle $ for the right Majorana in
Fig.~\ref{fig2}. We can use the lowest order perturbation theory to find the
hybridization energy of MBS provided that normalized solutions for the left
and right edges weakly overlap, i.e. $E^{\text{n(t)}}\approx|\left\langle
\psi_{\text{L}}\right|H\left|\psi_{\text{R}}\right\rangle |$ where the index
stands for the hybridization energy over the non-topological (topological)
region. For a T--N--T system in Fig.~\ref{fig2}, we obtain the hybridization
energy over the non-topological region:
\begin{equation}
\frac{E^{\text{n}}}{E_{0}^{\text{n}}}\approx e^{-\ell_{\text{n}}\mathfrak{Re}(\kappa_{2}^{\text{n}})}
\cos\left[\dfrac{\theta}{2}+\Phi_{0}+\ell_{\text{n}}\,\mathfrak{Im}(\kappa_{2}^{\text{n}})\right],\label{eq:HybNT}
\end{equation}
where
$\kappa_{2}^{n}=m^{*}/\hbar^2\bigl(i\alpha_{\text{so}}-i\sqrt{2(i\Delta+\mu)\hbar^2/m^{*}+\alpha_{\text{so}}^{2}}\bigr)$,
$E_{0}^{\text{n}}$ and $\Phi_{0}$ depend on parameters of the T and N-regions
and do not depend on $\ell_{\text{n}}$ and $\theta$ \cite{Note}. For the spin
current we obtain:
\[
j_{s}^{z}=\pm\dfrac{\partial E^{\text{n}}(\theta)}{\partial\theta},
\]
which shows that the torque $\partial E^{\text{n}}(\theta)/\partial\theta$
acting on the magnets in Fig.~\ref{fig1} is generated solely by the spin
current passing through the middle N-region \cite{Note}. For N--T--N system in Fig.~\ref{fig2}, we
obtain the hybridization energy over the topological region:
\begin{equation}
\dfrac{E^{\text{t}}}{E_{0}^{\text{t}}}\approx e^{-\ell_{\text{t}}\kappa_{2}^{\text{t}}}+\left|A_{0}
\right|e^{-\ell_{\text{t}}\mathfrak{Re}(\kappa_{1}^{\text{t}})}\cos\left[\arg A_{0}+\ell_{\text{t}}\,\mathfrak{Im}(\kappa_{1}^{\text{t}})\right],
\label{eq:HybT}
\end{equation}
where $\kappa_{1}^{\text{t}}$ and $\kappa_{2}^{\text{t}}$ are solutions of
equation
$\sqrt{B^{2}-[\kappa^{2}(\hbar^2/2m)^{2}+\mu]^{2}}=\Delta+\alpha_{\text{so}}\kappa$
satisfying the condition $\mathfrak{Re}(\kappa)>0$, $E_{0}^{\text{t}}$ and
$A_{0}$ depend on parameters of the T and N-regions and do not depend on
$\ell_{\text{t}}$ and $\theta$ \cite{Note}.

Fig.~\ref{fig2} shows the hybridization energies given by
Eqs.~(\ref{eq:HybNT}) and~(\ref{eq:HybT}) for parameters corresponding to an
InSb nanowire. We observe an exponential decay with separation and a
$4\pi$-periodic behavior with the relative angle of magnetic fields, which is
typical for TI edges \cite{Meng:PRB2012,Jiang:prb2013}. In addition, we find
an oscillatory behavior of energy as a function of separation between
the MBS. Such behavior has been predicted for MBS localized in vortices in
2D $p$-wave superconductors~\cite{Cheng:PRL2009}, yet it remained unclear whether these oscillations would persist over a non-topological region. In fact, the absence of oscillations was suggested in \cite{Sau:PRA2010} but the regime considered there corresponded to a fully depleted electron band.  

\textit{Numerical results.} --- We map the BdG Hamiltonian~(\ref{eq:BdG}) to
a tight binding model:
\begin{equation}
\begin{aligned}H= & \sum_{i,\sigma,\sigma'}\left[c_{i+1\sigma}^{\dagger}(-t_{0}\sigma_{0}+i\dfrac{\alpha_{i}}{2}\sigma_{z})_{\sigma\sigma'}c_{i\sigma'}+H.c.\right]\\
 & +\sum_{i,\sigma}(2t_{0}-\mu_{i})c_{i\sigma}^{\dagger}c_{i\sigma}+\sum_{i}(\widetilde{\Delta}_{i}c_{i\uparrow}^{\dagger}c_{i\downarrow}^{\dagger}+H.c.)\\
 & +\sum_{i}(\widetilde{B}_{i}c_{i\uparrow}^{\dagger}c_{i\downarrow}+H.c.)\,,
\end{aligned}
\label{eq:TBham}
\end{equation}
where we introduce complex parameters $\widetilde{\Delta}=\Delta e^{i\phi}$
and $\widetilde{B}=Be^{i\theta}$. In the long wavelength limit, the tight
binding model in Eq.~(\ref{eq:TBham}) can be reduced to Eq.~(\ref{eq:BdG})
with $t_{0}=\hbar^2/2m^{*}a^{2}$, $\alpha=\alpha_{\text{so}}/a$ where $a$ is
the lattice constant. For Fig.~\ref{fig2}, we use parameters consistent with
InSb quantum wires \cite{Mourik:Science2012}, i.e. $m^{*}=0.015m_{e}$,
$\alpha_{\text{so}}=0.2\,\textrm{eV\.\AA}$, $a=15\,${nm}, $\Delta=0.5\,$meV
and $g\mu_{B}=1.5\,$meV/T. The overall length of the wire corresponding to
Fig.~\ref{fig1} is $500$ sites. Results of our numerical diagonalization of
Hamiltonian (\ref{eq:TBham}) are presented in Fig.~\ref{fig2} by circles. We
observe perfect agreement with analytical Eqs.~(\ref{eq:HybNT}) and
(\ref{eq:HybT}) when T-regions are formed by uniform magnetic fields.

\begin{figure}
\includegraphics[clip,width=0.48\textwidth]{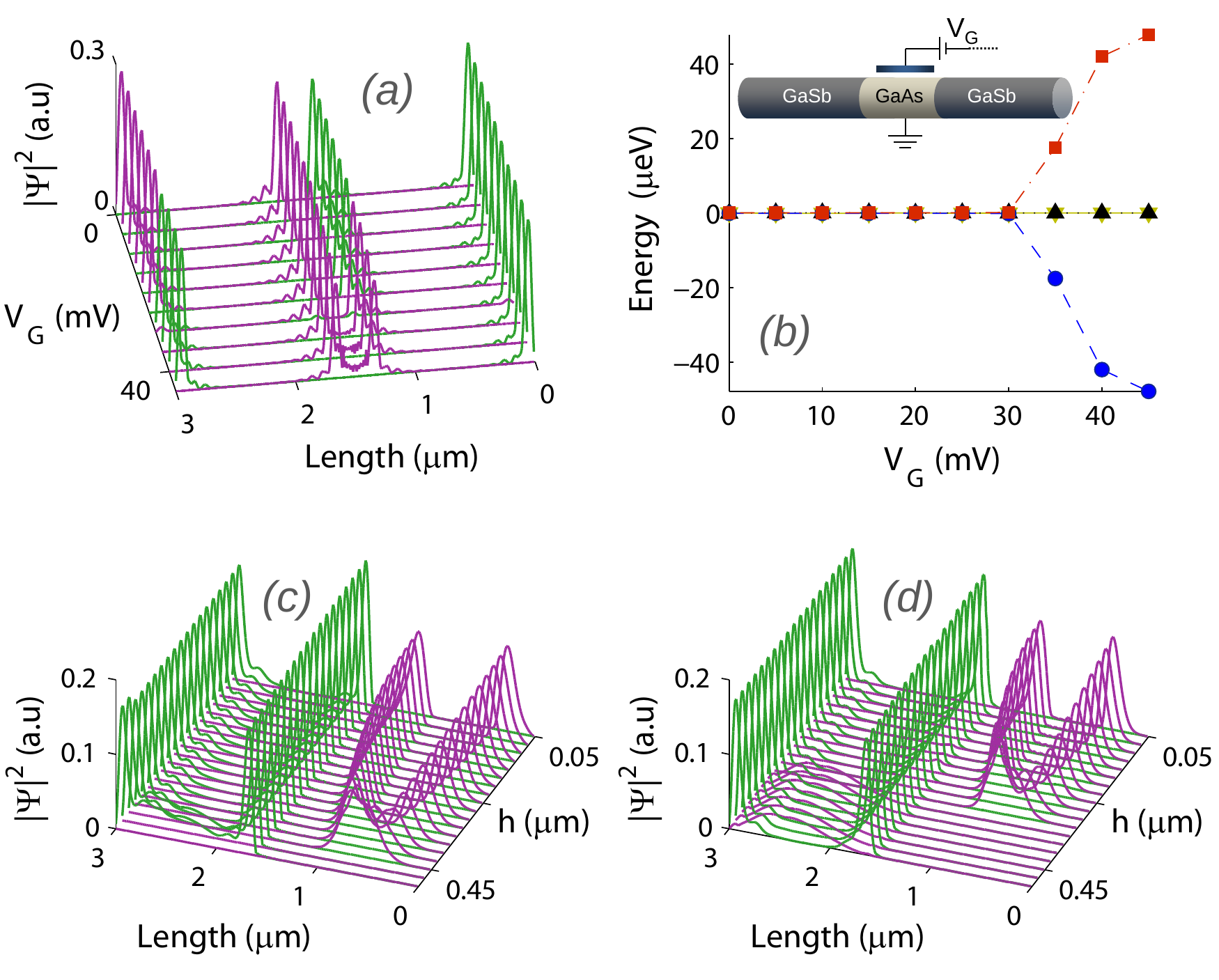} \caption{(Color online) (a) We plot edge modes as a function of position and
hetero-junction bias voltage $V_{G}$, and (b) the corresponding energies
as a function of $V_{G}$. We see hybridization at $V_{G}=30\, meV$.
(c) The wire is subjected to a constant magnetic field on one half
and a field of magnetic dipole at distance $h$ on the other half.
(d) The same but for a magnetic disk of radius $R=100~\text{nm}$
instead of dipole.}

\label{fig2-5}
\end{figure}

In Fig.~\ref{fig2-5}(a) and (b) we study the hybridization of MBS that are
defined by the modulation of the $g-$factor (by a factor of $30$) in GaSb-GaAs-GaSb type-II nanowire
heterostructures \cite{Guo2006,Ganjipour2011,De2007}.  Employment of such nanostructures can partially relax the requirement for the sharpness
of magnetic field profiles. Due to the bottom of the conduction band
mismatch a finite gate voltage is necessary in order to hybridize MBS. In
order to study the effect of non-uniform magnetic fields, in
Fig.~\ref{fig2-5}(c) the wire is subjected to a constant magnetic field on
one half and a field of magnetic dipole at distance $h$ on the other half.
Somewhat sharper MBS are formed when instead of a dipole we use a
perpendicularly magnetized thin disk in Fig.~\ref{fig2-5}(d). The magnetic
field decay length along the wire is defined by the distance $h$ between the
wire and the magnet in Fig.~\ref{fig1} which implies the requirement
$\ell_{\text{n}}\gtrsim h$.

\textit{Dissipative dynamics.} --- We suppose that the section of the wire
separating MBS $\gamma_{1}$ and $\gamma_{2}$ is sufficiently long (see
Fig.~\ref{fig1}). The effective low energy theory describing coupled dynamics
of MBS and a mechanical resonator can be described to the lowest order by
Hamiltonian in Eq.~(\ref{eq:Ham1}). Without loss of generality, we assume
that the electron parity in the wire is $1$ which defines the available
Hilbert space of two fermions $b_{1}=\gamma_{1}+i\gamma_{2}$ and
$b_{2}=\gamma_{3}+i\gamma_{4}$, i.e. $\alpha\left|1,0\right\rangle
+\beta\left|0,1\right\rangle $. By rewriting Eq.~(\ref{eq:Ham1}) through
fermionic operators $b_{1}$ and $b_{2}$, and expanding energies around
$\theta_{0}$, we arrive at the matrix Hamiltonian:
\begin{equation}
\mathcal{H}=\hbar\omega_{\text{r}}a^{\dagger}a+\left[\dfrac{E^{\text{n}}(\theta_{0})}{4}
+\dfrac{\partial E^{\text{n}}}{\partial\theta}\dfrac{\theta_{\text{zpf}}}{4}(a^{\dagger}+a)\right]\sigma_{x}+\dfrac{E^{\text{t}}}{4}\sigma_{z}\,,\label{eq:Rabi}
\end{equation}
where $E^{\text{n}}$ and $E^{\text{t}}$ are given by Eqs.~(\ref{eq:HybNT})
and (\ref{eq:HybT}). By tuning either $E^{\text{n}}(\theta_{0})/2$ or
$E^{\text{t}}/2$ to coincide with $\hbar\omega_{\text{r}}$ (see
Fig.~\ref{fig2}), we can achieve different regimes of Rabi oscillations. Note that when 
$\left\langle\psi_{\text{L}}\right|H\left|\psi_{\text{R}}\right\rangle $ is not pure imaginary we 
recover additional terms proportional to $\sigma_y$ in Eq.~(\ref{eq:Rabi}).
Here, we analyze the case in which $\hbar\omega_{\text{r}}=E^{\text{t}}/2$
and $E^{\text{n}}(\theta_{0})=0$. From Eq.~(\ref{eq:HybNT}) the coupling
strength (Rabi oscillations frequency) is
\[
g=\dfrac{1}{8}\theta_{\text{zpf}}E_{0}^{\text{n}}e^{-\ell_{\text{n}}\mathfrak{Re}(\kappa_{2}^{\text{n}})},
\]
which shows that by taking smaller $\ell_{\text{n}}$ we can increase the
coupling strength. The strong coupling regime can be realized when
$\omega_{\text{r}}/Q<g$ where $Q$ is the quality factor of the cantilever. A
pendulum based on single-walled carbon nanotube with an attached magnet of
the size $60\times40\times20\:\text{nm}^{3}$ can have
$K=3\times10^{-18}\text{Nm}$ per radian and
$I=4\times10^{-34}\text{kg}\cdot\text{\ensuremath{\text{m}^{2}}}$
\cite{Meyer:SEP2005}. If we take the corresponding
$\theta_{\text{zpf}}=5\times10^{-5}$, $\omega_{\text{r}}=5\text{MHz}$ and
$\ell_{\text{n}}=300\text{nm}$ (see Fig.~\ref{fig2}) we obtain $g=100\text{kHz}$
which is a strong coupling, e.g. a mechanical resonator with a resonant
frequency $\omega_{\text{r}}=5\text{MHz}$ will have to have relatively small
$Q>50$ in order to be in the strong coupling regime. In order to be able to
switch off interactions between the Majorana qubit and the resonator one can
use special $\ell_{\text{t}}$ points at which the hybridization energy is close
to zero (see Fig.~\ref{fig2}). In principle, $\ell_{\text{t}}$ can be controlled
by electrostatic gates \cite{Alicea:Nat2011} or supercurrents
\cite{Romito:PRB2012}.
\begin{figure}
\includegraphics[clip,width=0.5\textwidth]{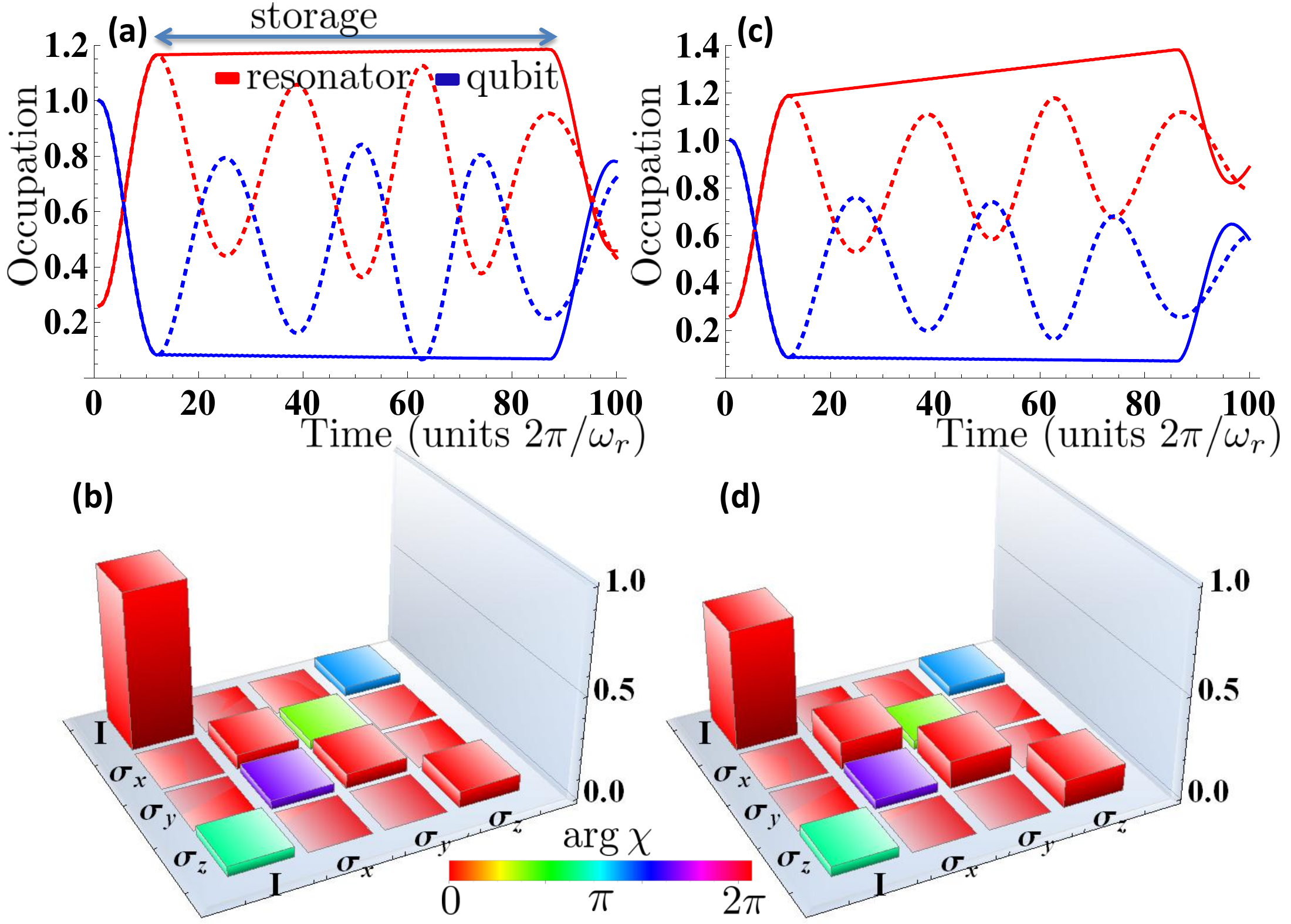} \caption{(Color online) (a) Rabi oscillations of a Majorana qubit coupled to
a mechanical resonator in Fig.~\ref{fig1}. (b) The quantum process
tomography of a process in which qubit state is transferred to the
resonator, then stored in the resonator while the systems are detuned,
and finally transferred back to the qubit. (c) and (d) same as (a)
and (b) but for a resonator with smaller quality factor.}

\label{fig4}
\end{figure}

The time dependent dissipative dynamics of the Hamiltonian (\ref{eq:Rabi})
can be adequately simulated using the Lindblad master equation \cite{Johansson2012}:
\begin{equation}
\dot{\rho}(t)=-\dfrac{i}{\hbar}\left[H(t),\rho\right]+\dfrac{1}{2}\sum_{k}
\left[\mathcal{L}_{k},\rho(t)\mathcal{L}_{k}^{\dagger}\right]
+\left[\mathcal{L}_{k}\rho(t),\mathcal{L}_{k}^{\dagger}\right],\label{eq:Lindblad}
\end{equation}
where we assume that all requirements on the environment for the validity of
this approximation apply. Here, $\mathcal{L}_{k}$ are Lindblad operators, in
particular $\mathcal{L}_{1}=\sqrt{1/T_{1}}\sigma_{-}$ and
$\mathcal{L}_{2}=\sqrt{1/T_{\phi}}\sigma_{+}\sigma_{-}$ correspond to the
majorana qubit coupling to the environment,
$\mathcal{L}_{3}=\sqrt{(\overline{n}_{\text{r}}+1)\omega_{\text{r}}/Q}a$ and
$\mathcal{L}_{4}=\sqrt{\overline{n}_{\text{r}}\omega_{\text{r}}/Q}a^{\dagger}$
correspond to the dissipation of the resonator where
$\overline{n}_{\text{r}}=[\exp(\omega_{\text{r}}/k_{B}T)-1]^{-1}$ and the
qubit lifetimes are given by $T_{1}$ and $1/T_{2}=1/2T_{1}+1/T_{\phi}$. The
Majorana qubit can decohere due to tunelling of fermions in the presence of  an external environment such as phonons, two-level systems,  classical noise~\cite{Goldstein:RPB2011}, as well as quasiparticle poisoning~\cite{Rainis:PRB2012}. As $T_{1}$ and $T_{2}$ times
can strongly depend on the concrete realization, in our simple analysis we
choose decoherence times that are order of the magnitude consistent with the
above mentioned mechanisms ($T_{1}=70\mu\text{s}$, $T_{2}=90\mu\text{s}$).

We present numerical solutions of Eq.~(\ref{eq:Lindblad}) for different
resonator quality factors, i.e. for $Q=10^{6}$ in Figs.~\ref{fig4}(a) and
(b), and for $Q=10^{5}$ in Figs.~\ref{fig4}(c) and (d). We assume the
resonator temperature $T=10\text{mK}$ and the initial occupation number
$\overline{n}_{\text{r}}=0.26$, e.g. as a result of sideband cooling
\cite{Teufel:jul2011}. Dotted lines represent the Rabi oscillations while the
bold lines represent the process in which the Majorana qubit is repeatedly
tuned in and out of resonance with the resonator. In such a process the qubit
state is transferred from the qubit to the resonator, then stored in the
resonator while the systems are detuned, and finally transferred back to the
qubit. We can completely describe the process of storage by the quantum
process tomography in which the final density matrix of the qubit is
described by the process matrix $\chi$, such that
$\rho_{\text{out}}=\sum\chi_{i,j}\sigma_{i}\rho_{\text{in}}\sigma_{j}$, here
$\sigma_{j}$ are Pauli matrices and $\sigma_{0}$ is the identity matrix. In
Figs.~\ref{fig4}(b) and (d) we plot the matrix $\chi$ where the two plots
correspond to fidelities $F=78\%$ and $F=60\%$, respectively.

\textit{Conclusions.} --- We demonstrated spin-current mediated resonant
coupling between a Majorana qubit and a mechanical resonator. The
coupling can manifest itself in a shift of the mechanical resonant
frequency, Rabi oscillations, coherent state transfer and Majorana
qubit/resonator entanglement. In addition, the spin-current mediated
coupling can facilitate both control of Majorana zero modes in a quantum
wire and transfer of quantum information between topological and conventional
qubits. The possibility to control the coupling and non-dissipative
spin currents in the spin-transistor type architecture paves the way
for applications in novel electronic devices. Our predictions can
be tested by employing the magnetic resonance force microscopy techniques.

We are grateful to Leonid Pryadko for multiple helpful discussions. AAK and AD
were supported in part by the U.S. Army Research Office under Grant
No.~W911NF-11-1-0027, and by the NSF under Grant No.~1018935. KS was
supported in part by the DARPA-QuEST program and by the NSF under Grant
DMR-0748925.

\section{Supplementary material}

In this supplementary material, we present more details on analytical
solutions for the edge states in topological (T) and non-topologial (N)
wires, and further apply these results to T--N--T and N--T--N setups.

\subsection{General solutions}

We consider the BdG Hamiltonian:
\begin{equation}
\begin{aligned}H= & k^{2}\hat{\tau}^{z}+uk\hat{\tau}^{z}\hat{\sigma}^{z}-\mu\hat{\tau}^{z}+\Delta(\cos\phi\hat{\tau}^{x}-\sin\phi\hat{\tau}^{y})\\
 & -b\hat{\sigma}^{z}+B(\cos\theta\hat{\sigma}^{x}-\sin\theta\hat{\sigma}^{y})\,,
\end{aligned}
\label{eq:BdG-s}
\end{equation}
where we use the Nambu spinor basis
$\Psi^{T}=(\psi_{\uparrow},\psi_{\downarrow},\psi_{\downarrow}^{\dagger},-\psi_{\uparrow}^{\dagger})$
and the Pauli matrices $\sigma^{i}$ and $\tau^{i}$ describe the spin and
particle-hole sectors, respectively. The Hamiltonian in Eq.~(\ref{eq:BdG}) is
written in dimensionless units where $u=\alpha_{\text{so}}/t_{0}a$ is the
dimensionless strength of spin-orbit interaction, $\mu$ is the chemical
potential, $\Delta e^{i\phi}$ is the superconducting pairing, $b$ is the
magnetic field along the $z$-direction and $B$ is the magnetic field in the
$xy$-plane. The energy unit is $t_{0}=\hbar/2m^{*}a^{2}$,
$\alpha_{\text{so}}$ is the Rashba spin-orbit coupling and the unit of length
$a$ is the lattice spacing for the tight binding representation of the
Hamiltonian. It is convenient to transform Eq.~(\ref{eq:BdG-s}) into the
following non-Hermitian form:
\[
G=V^{\dagger}U^{\dagger}HUV\tau^{z}\hat{\sigma}^{z},
\]
explicitly
\begin{equation}
G=k^{2}\hat{\sigma}^{z}+uk-(b\hat{\tau}^{z}+i\Delta\hat{\tau}^{x})-(\mu\hat{\sigma}^{z}+iB\hat{\sigma}^{x}).\label{eq:Gmatrix-s}
\end{equation}
The eigen solutions of the matrix $G$ correspond to the following
eigen values:
\begin{equation}
E_{i}=ku\pm\left[\sqrt{(\mu-k^{2})^{2}-B^{2}}\pm\sqrt{b^{2}-\Delta^{2}}\right].\label{eq:eigen-s}
\end{equation}
By taking the product of eigen values in Eq.~(\ref{eq:eigen-s}) one can
obtain the condition on a gapped phase, i.e. whenever there are real
solutions of equation:
\begin{equation}
\begin{aligned} & \left\{ k^{2}u^{2}-\left[\sqrt{(\mu-k^{2})^{2}-B^{2}}+\sqrt{b^{2}-\Delta^{2}}\right]\right\} \\
 & \times\left\{ k^{2}u^{2}-\left[\sqrt{(\mu-k^{2})^{2}-B^{2}}-\sqrt{b^{2}-\Delta^{2}}\right]\right\} =0,
\end{aligned}
\label{eq:gapped-s}
\end{equation}
the wire is in a gapless phase. When there are no real solutions of
Eq.~(\ref{eq:gapped-s}) the wire is in a gapped case. Here, we limit our
consideration by condition $\Delta^{2}>b^{2}$ in which case the Hamiltonian
(\ref{eq:BdG-s}) is always gapped with two phases, topological (T) for
$\Delta^{2}-b^{2}<B^{2}-\mu^{2}$ and non-topological (N) for
$\Delta^{2}-b^{2}>B^{2}-\mu^{2}$, being separated by a quantum phase
transition at $\Delta^{2}-b^{2}=B^{2}-\mu^{2}$.

In the gapped phase, we solve Eq.~(\ref{eq:gapped-s}) by substituting
$k=-i\kappa$ and finding $4-$component solution to the Hamiltonian
(\ref{eq:BdG-s}) in the form $\Psi(x)=e^{\kappa x}\Psi(\kappa)$ where we
arrive at four solutions with $\mathfrak{Re}(\kappa)>0$ and at four solutions
with $\mathfrak{Re}(\kappa)<0$. The general form of the corresponding
non-normalized eigen vectors becomes:
\[
\begin{aligned}\Psi(x)^{T}= & e^{\kappa x}\left(b\pm\sqrt{b^{2}-\Delta^{2}},i\Delta\right)\\
 & \otimes\left(\mu+\kappa^{2}\pm\sqrt{(\mu+\kappa^{2})^{2}-B^{2}},iB\right).
\end{aligned}
\]

\subsection{Solutions for topological and non-topological regions}

In order to avoid very complicated analytical expressions, in our discussion
we assume $b=0$. We arrive at four solutions for the topological region,
${\kappa_{-1}^{>}=\kappa_{1}^{\text{t}}}$,
${\kappa_{-2}^{>}=\kappa_{2}^{\text{t}}>0}$,
$\kappa_{-3}^{>}=\kappa_{1}^{\text{t}^\ast}$,
$\kappa_{+4}^{>}=\kappa_{3}^{\text{t}}>0$, for $\mathfrak{Re}(\kappa)>0$ and
 four solutions,
$\kappa_{-1}^{<}=-\kappa_{3}^{\text{t}}$,
$\kappa_{+2}^{<}=-\kappa_{1}^{\text{t}^\ast}$,
$\kappa_{+3}^{<}=-\kappa_{2}^{\text{t}}$,
$\kappa_{+4}^{<}=-\kappa_{1}^{\text{t}}$, for $\mathfrak{Re}(\kappa)<0$ where
$\kappa_{1}^{\text{t}}$, $\kappa_{1}^{\text{t}^\ast}$ and
$\kappa_{2}^{\text{t}}$ correspond to equations
$\pm\sqrt{B^{2}-(\kappa^{2}+\mu)^{2}}=\Delta+u\kappa$ and
$\kappa_{3}^{\text{t}}$ corresponds to equations
$\pm\sqrt{B^{2}-(\kappa^{2}+\mu)^{2}}=\Delta-u\kappa$. General $4-$component
unnormalized topological solutions take the form:
\begin{equation}
\Psi_{i}^{\text{t}}(x)=e^{x\kappa_{\mp i}^{>(<)}}VU\tau^{z}\sigma^{z}\left(\begin{array}{c}
\mp i(\mu+k^{2})-\Delta\mp u\kappa_{\mp i}^{>(<)}\\
B\\
-i(\mu+k^{2})\mp\Delta-u\kappa_{\mp i}^{>(<)}\\
B
\end{array}\right)\,,\label{eq:Tsol-s}
\end{equation}
where $V=e^{-i\frac{\pi}{4}\tau^{z}\sigma^{z}}$ and $U=e^{i\frac{\phi}{2}\tau^{z}}\otimes e^{i\frac{\theta}{2}\sigma^{z}}$.

In the non-topological region we assume that $B=0$, thus arriving at four
solutions,
$\kappa_{1}^{>}=\kappa_{1}^{n},\kappa_{2}^{>}=\kappa_{1}^{\text{n}\ast},\kappa_{3}^{>}=\kappa_{2}^{\text{n}},\kappa_{4}^{>}=\kappa_{2}^{\text{n}\ast}$,
for $\mathfrak{Re}(\kappa)>0$ and at four solutions,
$\kappa_{1}^{<}=-\kappa_{2}^{\text{n}\ast},\kappa_{2}^{<}=-\kappa_{2}^{\text{n}},\kappa_{3}^{<}=-\kappa_{1}^{\text{n}\ast},\kappa_{4}^{<}=-\kappa_{1}^{\text{n}}$,
for $\mathfrak{Re}(\kappa)<0$ where
$\kappa_{1}^{\text{n}}=iu/2+i\sqrt{\mu+u^{2}/4-i\Delta}$ and
$\kappa_{2}^{\text{n}}=iu/2-i\sqrt{\mu+u^{2}/4+i\Delta}$. The $4-$component
non-normalized solutions can be expressed in the following form:
\begin{equation}
\Psi_{i}^{\text{n}}(x)=e^{x\kappa_{i}^{<(>)}}VU\tau^{z}\sigma^{z}\Psi_{i}\,,\label{eq:NTsol-s}
\end{equation}
where $\Psi_{1}=(0,1,0,1)^{T}$, $\Psi_{2}=(1,0,1,0)^{T}$, $\Psi_{3}=(0,-1,0,1)^{T}$
and $\Psi_{4}=(-1,0,1,0)^{T}$.

\subsection{Hybridization of Majorana modes and spin currents in T--N--T wire}

We consider a semiconductor wire that has two infinite T-regions and a finite
N-region. We introduce parameters
$\{\Delta_{\text{l}},B_{\text{l}},\mu_{\text{l}},\theta_{\text{l}}\}$ for the
left T-regions, $\{\Delta_{\text{m}},\mu_{\text{m}}\}$ for the middle
N-region and
$\{\Delta_{\text{r}},B_{\text{r}},\mu_{\text{r}},\theta_{\text{r}}\}$ for the
right T-region (see Fig.~2, main text). The phase of superconducting pairing
is assumed constant (i.e. $\phi=0$) throughout the wire. The solutions in
Eqs.~(\ref{eq:Tsol-s}) and (\ref{eq:NTsol-s}) and their derivatives are
continuous at the boundary between the T and N-regions leading to unique
solution for the Majorana mode. We denote such solutions as
$\left|\psi_{\text{L}}\right\rangle
=e^{i\frac{\theta_{\text{l}}}{2}\hat{\sigma}^{z}}\left|\psi_{\text{L}}\right\rangle
$ for the left Majorana edge and as $\left|\psi_{\text{R}}\right\rangle
=e^{i\frac{\theta_{\text{r}}}{2}\hat{\sigma}^{z}}\left|\psi_{\text{R}}\right\rangle
$ for the right Majorana edge (see Fig.~2 in the main text) where it is
convenient to introduce solutions $\left|\psi_{\text{L}}^{0}\right\rangle $
and $\left|\psi_{\text{R}}^{0}\right\rangle $ corresponding to
$\theta_{\text{l}}=\theta_{\text{r}}=0$. When solutions for the left and
right edges weakly overlap we can find the hybridization energy of Majorana
modes and spin current at the N--T boundary by employing the lowest order
perturbation theory. For the hybridization energy we obtain:
\begin{align*}
E^{\text{n}}(\theta) & \approx\dfrac{\left|\left\langle \psi_{\text{L}}^{0}e^{-i\frac{\theta_{\text{l}}}{2}\hat{\sigma}^{z}}\right|H\left|e^{i\frac{\theta_{\text{r}}}{2}\hat{\sigma}^{z}}\psi_{\text{R}}^{0}\right\rangle \right|}{\sqrt{\left\langle \psi_{\text{L}}^{0}|\psi_{\text{L}}^{0}\right\rangle \left\langle \psi_{\text{R}}^{0}|\psi_{\text{R}}^{0}\right\rangle }},
\end{align*}
and for spin current we obtain:\begin{widetext}

\[
j_{s}^{z}(x_{\text{R}})=\dfrac{\mathfrak{Re}\left\{ \left[\psi_{\text{L}}^{\dagger}(x_{\text{R}})\mp i\psi_{\text{R}}^{\dagger}(x_{\text{R}})\right]\hat{\sigma}^{z}\hat{\upsilon}\Bigl[\psi_{\text{L}}(x_{\text{R}})\pm i\psi_{\text{R}}(x_{\text{R}})\Bigr]\right\} }{2\sqrt{\left\langle \psi_{\text{L}}^{0}|\psi_{\text{L}}^{0}\right\rangle \left\langle \psi_{\text{R}}^{0}|\psi_{\text{R}}^{0}\right\rangle }},
\]
where $x_{\text{L}}$ and $x_{\text{R}}$ are positions of the edge states,
$\hat{\upsilon}=\partial\hat{H}/\partial p$,
$\theta=\theta_{\text{r}}-\theta_{\text{l}}$ and
$j_{s}^{z}(x_{\text{L}})=j_{s}^{z}(x_{\text{R}})$. Explicitly, we have
\[
\left\langle \psi_{\text{L}}^{0}e^{-i\frac{\theta_{\text{l}}}{2}\hat{\sigma}^{z}}\right|H\left|e^{i\frac{\theta_{\text{r}}}{2}\hat{\sigma}^{z}}\psi_{\text{R}}^{0}\right\rangle =2e^{-2\ell_{\text{n}}\mathfrak{Re}(\kappa_{2}^{\text{m}})}\left[L_{\text{R}}^{1}R_{\text{L}}^{2^{\ast}}(2\kappa_{2}^{\text{m}^{\ast}}+iu)e^{\ell_{\text{n}}\kappa_{2}^{\text{m}}+\frac{i\theta}{2}}+L_{\text{R}}^{2}R_{\text{L}}^{1^{\ast}}(2\kappa_{2}^{\text{m}}-iu)e^{\ell_{\text{n}}\kappa_{2}^{\text{m}^{\ast}}-\frac{i\theta}{2}}\right],
\]
and spin current becomes:
\[
j_{s}^{z}=\pm\dfrac{\partial E^{\text{n}}(\theta)}{\partial\theta},
\]
\end{widetext}which corresponds to the formula for the hybridization
energy over the non-topological region in the main text:
\begin{equation}
\dfrac{E^{\text{n}}}{E_{0}^{\text{n}}}\approx e^{-\ell_{\text{n}}\mathfrak{Re}(\kappa_{2}^{\text{m}})}\cos\left[\dfrac{\theta}{2}+\Phi_{0}+L\mathfrak{Im}(\kappa_{2}^{\text{m}})\right],\label{eq:HybNT-s}
\end{equation}
 with
\begin{align*}
\Phi_{0} & =\arg\left[2\kappa_{2}^{\text{m}}-iu\right]+\dfrac{1}{2}\arg\left[L_{\text{R}}^{2}R_{\text{L}}^{1^{\ast}}/(L_{\text{R}}^{1}R_{\text{L}}^{2^{\ast}})\right].
\end{align*}
Here $\left|\psi_{\text{L}}^{0}\right\rangle $ and
$\left|\psi_{\text{R}}^{0}\right\rangle $ can be written as

\begin{align}
\left|\psi_{\text{L}}^{0}\right\rangle  & =\begin{cases}
\sum_{i=1}^{4}L_{\text{L}}^{i}\Psi_{i}^{\text{t}}(x); & x<x_{\text{L}}\\
\sum_{i=1}^{4}R_{\text{L}}^{i}\Psi_{i}^{\text{n}}(x); & x>x_{\text{L}}
\end{cases},\label{eq:solutions0-s}\\
\left|\psi_{\text{R}}^{0}\right\rangle  & =\begin{cases}
\sum_{i=1}^{4}L_{\text{R}}^{i}\Psi_{i}^{\text{n}}(x); & x<x_{\text{R}}\\
\sum_{i=1}^{4}R_{\text{R}}^{i}\Psi_{i}^{\text{t}}(x); & x>x_{\text{R}}
\end{cases},\label{eq:solutions-s}
\end{align}
where $x_{\text{L}}$ and $x_{\text{R}}$ are positions of the edge states and
\begin{widetext}

\begin{equation}
\begin{array}{c}
\begin{aligned}L_{\text{L}}^{1}= & \frac{(i\ensuremath{\Delta_{\text{l}}}-\ensuremath{\mu_{\text{l}}})(\kappa_{2}^{\text{m}}-\kappa_{2}^{\text{m}^{\ast}})-\kappa_{1}^{\text{l}^{\ast}}\left(\kappa_{2}^{\text{m}^{\ast}}+\kappa_{2}^{\text{l}}\right)(u+i\kappa_{2}^{\text{l}}+i\kappa_{2}^{\text{m}})-i(\kappa_{1}^{\text{l}^{\ast}})^{2}(\kappa_{2}^{\text{m}^{\ast}}+\kappa_{2}^{\text{l}})-\kappa_{2}^{\text{m}^{\ast}}(\kappa_{2}^{\text{l}}+\kappa_{2}^{\text{m}})(u+i\kappa_{2}^{\text{l}})}{(\kappa_{1}^{\text{l}}-\kappa_{1}^{\text{l}^{\ast}})(\kappa_{1}^{l}-\kappa_{2}^{\text{l}})\left(\kappa_{1}^{\text{l}^{\ast}}+\kappa_{1}^{\text{l}}+\kappa_{2}^{\text{l}}+\kappa_{2}^{\text{m}}-iu\right)},\\
L_{\text{L}}^{2}= & \frac{-\kappa_{1}^{\text{l}^{\ast}2}(\kappa_{2}^{\text{m}^{\ast}}+\kappa_{1}^{\text{l}})+i\kappa_{1}^{\text{l}^{\ast}}\left(\kappa_{2}^{\text{m}^{\ast}}+\kappa_{1}^{\text{l}}\right)(u+i\kappa_{1}^{\text{l}}+i\kappa_{2}^{\text{m}})+\kappa_{2}^{\text{m}^{\ast}}(\kappa_{1}^{\text{l}}+\kappa_{2}^{\text{m}})(iu-\kappa_{1}^{\text{l}})+(\Delta_{\text{\text{l}}}+i\mu_{\text{l}})(\kappa_{2}^{\text{m}}-\kappa_{2}^{\text{m}^{\ast}})}{(\kappa_{1}^{\text{l}}-\kappa_{2}^{\text{l}})\left(\kappa_{2}^{\text{l}}-\kappa_{1}^{\text{l}^{\ast}}\right)\left(\kappa_{1}^{\text{l}^{\ast}}+\kappa_{1}^{\text{l}}+\kappa_{2}^{\text{l}}+\kappa_{2}^{m}-iu\right)},\\
L_{\text{L}}^{3}= & \frac{i\kappa_{2}^{\text{m}^{\ast}}\left[(\kappa_{1}^{\text{l}})^{2}+\kappa_{2}^{\text{m}}(\kappa_{1}^{\text{l}}+\kappa_{2}^{\text{l}})+(\kappa_{2}^{\text{l}})^{2}-iu(\kappa_{1}^{\text{l}}+\kappa_{2}^{\text{l}}+\kappa_{2}^{\text{m}})\right]+2(\mu_{\text{l}}-i\Delta_{\text{l}})\mathfrak{Im}(\kappa_{2}^{\text{m}})+i\kappa_{1}^{\text{l}}\kappa_{2}^{\text{l}}(\kappa_{1}^{\text{l}}+\kappa_{2}^{\text{l}}+2\mathfrak{Re}(\kappa_{2}^{\text{m}})-iu)}{2\mathfrak{Im}(\kappa_{1}^{\text{l}})\left(\kappa_{1}^{\text{l}^{\ast}}-\kappa_{2}^{\text{l}}\right)\left(\kappa_{1}^{\text{l}^{\ast}}+\kappa_{1}^{\text{l}}+\kappa_{2}^{\text{l}}+\kappa_{2}^{\text{m}}-iu\right)},\\
R_{\text{L}}^{2}= & \frac{\left(-|\kappa_{1}^{\text{l}}|^{2}\left(\kappa_{2}^{\text{m}^{\ast}}+\kappa_{2}^{\text{l}}\right)+\kappa_{2}^{\text{m}^{\ast}}\left(-i\Delta_{\text{l}}+\mu_{\text{l}}+i(\kappa_{1}^{\text{l}}+\kappa_{1}^{\text{l}^{\ast}})(u+i\kappa_{2}^{\text{l}})+u^{2}+i\kappa_{2}^{\text{l}}u\right)-(\Delta_{\text{l}}+i\mu_{\text{l}})(i\kappa_{2}^{\text{l}}+i(\kappa_{1}^\text{{l}}+\kappa_{1}^{\text{l}^{\ast}})+u)\right)}{B_{\text{l}}(i(\kappa_{2}^{\text{l}}+\kappa_{2}^{\text{m}})+i(\kappa_{1}^{\text{l}}+\kappa_{1}^{\text{l}^{\ast}})+u)},\\
L_{\text{L}}^{4}= & R_{\text{L}}^{3}=R_{\text{L}}^{4}=L_{\text{R}}^{3}=L_{\text{R}}^{4}=R_{\text{R}}^{4}=0,\, R_{\text{L}}^{1}=L_{\text{R}}^{2}=1.
\end{aligned}
\end{array}\label{eq:formula}
\end{equation}

\end{widetext}The coefficients $L_{\text{R}}^{1}$, $R_{\text{R}}^{1}$, $R_{\text{R}}^{2}$
and $R_{\text{R}}^{3}$ can be obtained from $R_{\text{L}}^{2}$,
$L_{\text{L}}^{1}$, $L_{\text{L}}^{2}$ and $L_{\text{L}}^{3}$, respectively,
by replacement $\kappa_{1}^{\text{l}}\rightarrow-\kappa_{1}^{\text{r}}$,
$\kappa_{2}^{\text{l}}\rightarrow-\kappa_{2}^{\text{r}}$,
$\kappa_{2}^{\text{m}}\rightarrow-\kappa_{2}^{\text{m}^{\ast}}$,
$\mu_{\text{l}}\rightarrow\mu_{\text{r}}$, $B_{\text{l}}\rightarrow
B_{\text{r}}$ and $\Delta_{\text{l}}\rightarrow\Delta_{\text{r}}$.

\subsection{Hybridization of Majorana modes in N--T--N wire}

We consider a semiconductor wire that has two infinite N-regions and a finite
T-region. We introduce parameters $\{\Delta_{\text{l}},\mu_{\text{l}}\}$ for
the left N-regions,
$\{\Delta_{\text{m}},B_{\text{m}},\mu_{\text{m}},\theta_{\text{m}}\}$ for the
middle T-region and $\{\Delta_{\text{r}},\mu_{\text{r}}\}$ for the right
N-region (see Fig.~2, main text). The phase of superconducting pairing is
assumed constant (i.e. $\phi=0$) throughout the wire. The solutions in
Eqs.~(\ref{eq:Tsol-s}) and (\ref{eq:NTsol-s}) and their derivatives are
continuous at the boundary between the T and N-regions leading to unique
solution for the Majorana mode. We denote such solutions as
$\left|\psi_{\text{L}}\right\rangle $ for the left Majorana edge and as
$\left|\psi_{\text{R}}\right\rangle $ for the right Majorana edge (see Fig.~2
in the main text). Here, the angle $\theta_{\text{m}}$ does not play any
role. When solutions for the left and right edges weakly overlap we can find
the hybridization energy of Majorana modes by employing the lowest order
perturbation theory , i.e.
\begin{align*}
E^{\text{n}} & \approx\dfrac{\left|\left\langle \psi_{\text{L}}\right|H\left|\psi_{\text{R}}\right\rangle \right|}{\sqrt{\left\langle \psi_{\text{L}}|\psi_{\text{L}}\right\rangle \left\langle \psi_{\text{R}}|\psi_{\text{R}}\right\rangle }},
\end{align*}
Explicitly, we obtain\begin{widetext}
\[
\begin{aligned}\left\langle \psi_{\text{L}}\right|H\left|\psi_{\text{R}}\right\rangle = & \dfrac{2}{B_{\text{m}}}e^{-L_{\text{t}}\kappa_{1}^{\text{m}}}R_{\text{R}}^{3^{\ast}}\left(B_{\text{m}}R_{\text{L}}^{1}\left(-\kappa_{2}^{\text{r}^{\ast}}+\kappa_{1}^{\text{m}}-iu\right)+R_{\text{L}}^{2}(\kappa_{1}^{\text{m}}-\kappa_{2}^{\text{r}}+iu)\left(\Delta_{\text{m}}+i\left((\kappa_{1}^{\text{m}})^{2}+\mu_{\text{m}}\right)+\kappa_{1}^{\text{m}}u\right)\right)\\
 & +\dfrac{2}{B_{\text{m}}}e^{-L_{\text{t}}\kappa_{1}^{\text{m}^{\ast}}}R_{\text{R}}^{1^{\ast}}\left(B_{\text{m}}R_{\text{L}}^{1}\left(\kappa_{1}^{\text{m}^{\ast}}-\kappa_{2}^{\text{r}^{\ast}}-iu\right)+R_{\text{L}}^{2}\left(\kappa_{1}^{\text{m}^{\ast}}-\kappa_{2}^{\text{r}}+iu\right)\left(\kappa_{1}^{\text{m}^{\ast}}\left(u+i\kappa_{1}^{\text{m}^{\ast}}\right)+\Delta_{\text{m}}+i\mu_{\text{m}}\right)\right)\\
 & +\dfrac{2}{B_{\text{m}}}e^{-L_{\text{t}}\kappa_{2}^{\text{m}}}R_{\text{R}}^{2^{\ast}}\left(B_{\text{m}}R_{\text{L}}^{1}\left(-\kappa_{2}^{\text{r}^{\ast}}+\kappa_{2}^{\text{m}}-iu\right)+R_{\text{L}}^{2}(\kappa_{2}^{\text{m}}-\kappa_{2}^{\text{r}}+iu)\left(\Delta_{\text{m}}+i\left((\kappa_{2}^{\text{m}})^{2}+\mu_{\text{m}}\right)+\kappa_{2}^{\text{m}}u\right)\right),
\end{aligned}
\]
which corresponds to the formula for the hybridization energy over
the topological region in the main text:
\begin{equation}
\dfrac{E^{\text{t}}}{E_{0}^{\text{t}}}\approx e^{-\ell_{\text{t}}\kappa_{2}^{\text{m}}}+\left|A_{0}\right|e^{-\ell_{\text{t}}\mathfrak{Re}(\kappa_{1}^{\text{m}})}\cos\left[\arg A_{0}+\ell_{{\text{t}}}\mathfrak{Im}(\kappa_{1}^{\text{m}})\right],\label{eq:HybT-s}
\end{equation}
 with
\[
A_{0}=\dfrac{R_{\text{R}}^{1^{\ast}}\left(B_{\text{m}}R_{\text{L}}^{1}\left(\kappa_{1}^{\text{m}^{\ast}}-\kappa_{2}^{\text{r}^{\ast}}-iu\right)+R_{\text{L}}^{2}\left(\kappa_{1}^{\text{m}^{\ast}}-\kappa_{2}^{\text{r}}+iu\right)\left(\kappa_{1}^{\text{m}^{\ast}}\left(u+i\kappa_{1}^{\text{m}^{\ast}}\right)+\Delta_{\text{m}}+i\mu_{\text{m}}\right)\right)}{R_{\text{R}}^{2^{\ast}}\left(B_{\text{m}}R_{\text{L}}^{1}\left(-\kappa_{2}^{\text{r}^{\ast}}+\kappa_{2}^{\text{m}}-iu\right)+R_{\text{L}}^{2}(\kappa_{2}^{\text{m}}-\kappa_{2}^{\text{r}}+iu)\left(\Delta_{\text{m}}+i\left((\kappa_{2}^{\text{m}})^{2}+\mu_{\text{m}}\right)+\kappa_{2}^{\text{m}}u\right)\right)}.
\]
\end{widetext}Here we can take $\left|\psi_{\text{L}}\right\rangle =\left|\psi_{\text{R}}^{0}\right\rangle $
and $\left|\psi_{\text{R}}\right\rangle
=\left|\psi_{\text{L}}^{0}\right\rangle $ in Eqs.~(\ref{eq:solutions0-s}) and
(\ref{eq:solutions-s}) after replacement $x_{\text{L}}\leftrightarrow
x_{\text{R}}$ since we can use localized solutions found in the previous
section, i.e. we can use Eq.~(\ref{eq:formula}) for coefficients
$R_{\text{L}}^{i}$ and $L_{\text{L}}^{i}$ after replacement of indices
``m''$\rightarrow$``r'' and ``l''$\rightarrow$``m'' and we can use
coefficients $R_{\text{R}}^{i}$ and $L_{\text{R}}^{i}$ after replacement of
indices ``m''$\rightarrow$``l'' and ``r''$\rightarrow$``m''.

\bibliographystyle{apsrev}

\bibliographystyle{apsrev}
\bibliography{Mwire}

\end{document}